\documentclass[%
reprint,
superscriptaddress,
amsmath,amssymb,
aps,
prl,
]{revtex4-2}

\usepackage{graphicx}
\usepackage{dcolumn}
\usepackage{bm}
\usepackage{amsmath}
\usepackage[utf8]{inputenc}
\usepackage[T1]{fontenc}
\usepackage{booktabs, array, mathptmx, float, tabularx, booktabs, lipsum, amsmath,multirow}
\usepackage{siunitx, xcolor}
\usepackage[version=4]{mhchem}
\usepackage[colorlinks,colorlinks=true,
linkcolor=blue,
filecolor=blue,      
urlcolor=blue,
citecolor=blue,]{hyperref}

\newcommand{\mold}[1]{\iffalse{#1}\fi}
\draft 
\begin{document}

\preprint{APS/123-QED}

\title{Ultrafast High Energy Electron Radiography for Electromagnetic Field Diagnosis} 

\author{J. H. Xiao}
\affiliation{Department of Engineering Physics, Tsinghua University, Beijing 100084, China}
\affiliation{Key Laboratory of Particle and Radiation Imaging (Tsinghua University), Ministry of Education, Beijing 100084, China}
\author{Y. C. Du}
\email{dych@tsinghua.edu.cn}
\affiliation{Department of Engineering Physics, Tsinghua University, Beijing 100084, China}
\affiliation{Key Laboratory of Particle and Radiation Imaging (Tsinghua University), Ministry of Education, Beijing 100084, China}
\author{H. Q. Li}
\affiliation{Department of Engineering Physics, Tsinghua University, Beijing 100084, China}
\affiliation{State Key Laboratory of Intense Pulsed Radiation Simulation and Effect, Northwest Institute of Nuclear Technology, Xi’an 710024, China}
\author{Y. T. Zhao}
\affiliation{Xi’an Jiaotong University, Xi’an 710049, China}
\author{L. Sheng}
\affiliation{State Key Laboratory of Intense Pulsed Radiation Simulation and Effect, Northwest Institute of Nuclear Technology, Xi’an 710024, China}

\begin{abstract}
This letter proposes a new method based on ultrafast high energy electron radiography to diagnose transient electromagnetic field. For the traditional methods, large scattering from matter will increase the uncertainty of measurement, but our method still works in that case. To verify its feasibility, a $ 50MeV $ electron radiography beamline is designed and optimized, and preliminary simulation of diagnosing a circular magnetic field ranging from $ 170T*\mu m $ to $ \sim 600T*\mu m$ has been done. The simulation results indicate that this method can achieve point-by-point measurement of field strength. By destroying the angle symmetry of incident beams, the field direction can also be determined. Combined with the advantages of electron beams, ultrafast high energy electron radiography is very suitable for transient electromagnetic field diagnosis.

\end{abstract}

\maketitle

Transient electromagnetic field ($E/B$) plays an important role in advanced accelerator physics\cite{PhysRevLett.43.267,article2}, plasma physics\cite{PhysRevLett.78.254,article4,book1}, inertial confinement fusion (ICF)\cite{PhysRevLett.118.155001,PhysRevLett.103.215004}, high energy density physics\cite{650905}, experimental astrophysics, etc\cite{PhysRevLett.97.255001,doi:10.1063/1.3694119}. The $E/B$ fields in the above scenarios are difficult to diagnose due to their ultrafast evolving and submillimeter space scale. Thus, ultrafast diagnostics with high spatial resolution is required. Earlier works like C.k.Li et al.\cite{PhysRevLett.97.135003,PhysRevLett.100.225001,Rygg1223} and Schumaker et al.\cite{PhysRevLett.110.015003} successfully demonstrated charged particles like proton and electron adopted to make shadow image. Experimental and simulation results show that charged particle lens radiography technology has higher spatial resolution\cite{750705,doi:10.1063/1.4939822,doi:10.1063/1.4971560,PhysRevApplied.11.034068}. Based on this, relativistic electron beam with ultrashort bunch length might be a suitable probe to diagnose transient electromagnetic field due to its lower magnetic rigidity and high temporal resolution. Previously, the application of  high energy electron lens radiography (HEELR) in fluid diagnosis has been paid more attention to\cite{doi:10.1063/1.5109855,doi:10.1063/1.5011198,Xiao_2018,zhao}. Recently, HEELR has been proposed as an approach to observe ultrafast field evolution in a program\cite{article25}. In this letter, a new design is introduced to get the field direction and strength simultaneously. 

Above all, three approximations should be satisfied to ensure the reliability of this diagnosis. Firstly, the electric field in plasma can be taken as electrostatic field and there is no energy change from the field when electrons pass through the system. Secondly, the ion fluid evolution is relatively slow so the unevenness of ion distribution can be neglected. Lastly, the interactions of electrons with matter and electromagnetic field are independent, the matter makes electrons scattered randomly and the electromagnetic field makes electrons deflected to a specific direction.

The angle distribution of electrons after interacting with matter can be described by formula(\ref{Formula1})
\begin{equation}
	\label{Formula1}
	F(\xi)=\dfrac{N(t,\xi)}{N_0}=\dfrac{1}{\xi_0(t)\sqrt{2\pi}}exp(-\dfrac{1}{2}(\dfrac{\xi}{\xi_0(t)})^2)
\end{equation}
where
\begin{equation}
	\xi_0(t)=\dfrac{13.6MeV}{\beta cp}\sqrt{t}(1+0.038ln(t))\nonumber
\end{equation}
$\xi$ is the scattering angle, $t=X/X_0$ is the relative thickness of target, $X$ is thickness of target, $X_0$ is the radiation length of target material, $p$ is the electron momentum.
Thus after penetrating a uniform matter, electrons will have a specific angle distribution with circular symmetry in transverse direction. The probability density of the distribution can be represented by $j_{\bm{\xi}}(x',y')$ in $ x'-y' $ coordinate system, and it can be converted to $ f_{\bm{\xi}}(\rho',\varphi') $ in polar coordinate, where $ \rho'$ is the radius and $\varphi'$ is the polar angle. The footnote $\bm{\xi}$ indicates the distribution is of scattering with matters. Considering the symmetry in transverse direction, $f_{\bm{\xi}}$ is only related to $\rho'$ in fact. So $ f_{\bm{\xi}}(\rho',\varphi') $  can by noted as  $ f_{\bm{\xi}}(\rho') $ briefly. This also implies that the angle distribution of incident electrons can be controlled by a scattering target to meet the preferred diagnostic requirement.

On the other hand, when electromagnetic field exists in the target, the electron will be deflected. The deflection angle from magnetic field is decided by:
\begin{equation}\label{Formula2}
	\begin{aligned}		
		\bm{\theta}&=\int{\bm{\omega}}dt\approx\int{\dfrac{e\textbf{B}_\perp(x,y,z)}{{\gamma}m_0}\times{\dfrac{\textbf{\textit{v}}}{|\textbf{\textit{v}}|}}}dt\\
		&=\int{\dfrac{e\textbf{B}_\perp(x,y,z)}{{\gamma}m_0\textit{v}}}\times{d\bm{z}}=\int{\dfrac{e\textbf{B}_\perp(x,y,z)}{p}}\times{d\bm{z}}
	\end{aligned}
\end{equation}
Here $ m_0 $ refers to the electron rest mass, $ \textbf{B}_\perp $ is the transverse component of magnetic field, z is the thickness of magnetic field.
Similarly, the deflection angle electric field contributes can be described as:
\begin{equation}\label{Formula3}
	\begin{aligned}	
		\bm{\theta}=\int{\frac{\textbf{F}_\perp{dt}}{p}}&=\int{\frac{{\textbf{E}_\perp(x,y,z){e}\frac{dz}{c}}}{p}}\\
		&=\int{\frac{\textbf{E}_\perp(x,y,z){e}}{cp}}{dz}
	\end{aligned}
\end{equation}
where $ \textbf{E}_\perp $ is the transverse component of electric field.
Consequently, the probability density distribution of electron's angle after target which includes both matters and $E/B$ field can be denoted by $f_{{\bm{\xi}}+{\bm{\theta}}}(\rho',\varphi')$ (or $j_{\bm{\xi}+\bm{\theta}}(x',y')$), the footnote ${\bm{\theta}}$ indicates the deflection from $E/B$ field is considered. According to the third approximation mentioned earlier in this letter, $f_{{\bm{\xi}}+{\bm{\theta}}}(\rho',\varphi')$ is $f_{{\bm{\xi}}}(\rho')$ with a shift in coordinates actually. The shift distance $\rho'_{\bm{\theta}}$ is related to the field strength and the shift direction $\varphi'_{\bm{\theta}}$ is connected with the field direction. \mold{It's worth mentioning that $\rho'_{{\bm{\xi}}+{\bm{\theta}}}$ is not $\rho'_{\bm{\xi}}+\rho'_\theta$ and $\phi'_{{\bm{\xi}}+{\bm{\theta}}}$ is not $\varphi'_{\bm{\xi}}+\varphi'_{\bm{\theta}}$.}

The above analysis demonstrates that the angle distribution of penetrating electrons contains the information including areal density and $E/B$ field of diagnosed target.  When the thin target approximation ($\rho'_{\bm{\theta}}>>\rho'_{\bm{\xi}}$) is satisfied, shadow image with charged particles has achieved good results. The shadow image diagnostics includes dark spot mode\cite{doi:10.1063/1.3491994,doi:10.1063/1.3380846} and beam meshing mode\cite{Chen14479}. When charged particles propagate through the field area, they will be deflected. Accordingly, the area with field will appear to be dim in the image. Up to a point, the dark spot image can mark the field information like the strength or transmission speed. On this basis, beam meshing experiment can get the field direction. This letter proposes a method to detect the field strength and direction in one shot when $\rho'_{\bm{\xi}}$ is analogous to $\rho'_{\bm{\theta}}$.

\begin{figure}[htp]
	\includegraphics[width=3in]{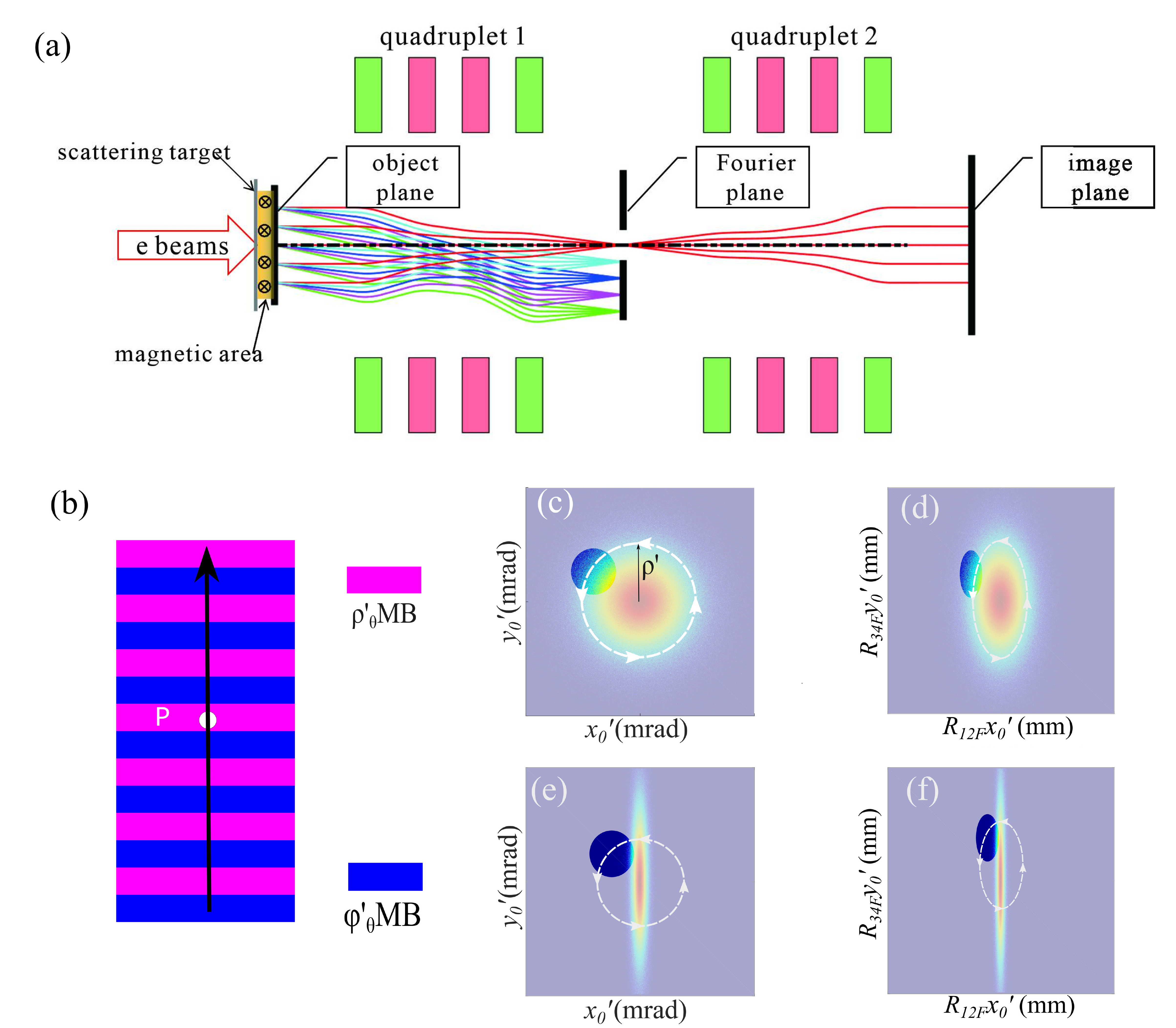}%
	\caption{\label{1}The schematic diagram of DTD based on HEELR. (a) The schematic of point-to-point imaging beamline with an aperture at Fourier plane to make angle selection. (b) Two kinds of target area, with(pink) and without(blue) destroying the circular symmetry of angle distribution of incident beams. (c) and (e) illustrate the angle distributions of beams penetrating pink area and blue area in (b) respectively. (d) and (f) are the corresponding beam patterns at the Fourier plane. Frosted part indicates electrons are blocked by aperture.}
\end{figure}

HEELR can achieve angle selection and refocusing for the penetrating electrons. For monoenergetic electron beams, its transportation in beamline can be expressed by the following matrix equation:
\begin{equation}
	\left[
	\begin{array}{cccc}
		x_{i}\\
		x_{i}^{'}\\
		y_{i}\\
		y_{i}^{'}\\
	\end{array}
	\right ]
	=
	\left[
	\begin{array}{cccc}
		R_{11}& R_{12} &0 & 0\\
		R_{21}&R_{22} &0 &0\\
		0&0 &R_{33} &R_{34}\\
		0&0 &R_{43} &R_{44}\\
	\end{array}
	\right ]
	\left[
	\begin{array}{cccc}
		x_{0}\\
		x_{0}^{'}\\
		y_{0}\\
		y_{0}^{'}
	\end{array}
	\right ]
\end{equation}
where $ x_0 $, $ y_0 $ and $ x_0' $, $ y_0' $ are the position and angle of electron at the object back plane, $ x_i $, $ y_i $ and $ x_i' $, $ y_i' $ are the corresponding coordinates at the image plane,  $ R_{ij} $ is the transfer matrix factor of the beamline. For the entire image system, $R_{12}=R_{34}=0$ means the position of electron at the image plane is independent from the initial angle so we can obtain the point-to-point image, as shown in Fig.\ref{1}.a. Another important plane named Fourier plane exists between the object plane and the image plane. At this plane, $R_{11F} $ and $ R_{33F} $ are 0, the subscript $F$ indicates from the object plane to the Fourier plane. If the Fourier plane of x-axis and y-axis are coincident along z-axis by optimization, the following equation can be satisfied:
\begin{equation}
	\left[
	\begin{array}{cc}
		x_{F}\\
		y_{F}\\
	\end{array}
	\right ]
	=
	\left[
	\begin{array}{cc}
		R_{12F} &0\\
		0&R_{34F}\\
	\end{array}
	\right ]
	\left[
	\begin{array}{cc}
		x_{0}^{'}\\
		y_{0}^{'}
	\end{array}
	\right ]
\end{equation}
This implies that the electron's angle information at the object plane $j(x'_0,y'_0)$ has been translated into position distribution $g(x_F,y_F)$ at this plane. Apparently, an aperture at this plane can achieve angle selection for propagating electrons, as shown in Fig.\ref{1}.a.
\begin{figure*}[htp]
	\centering
	\includegraphics[width=6in]{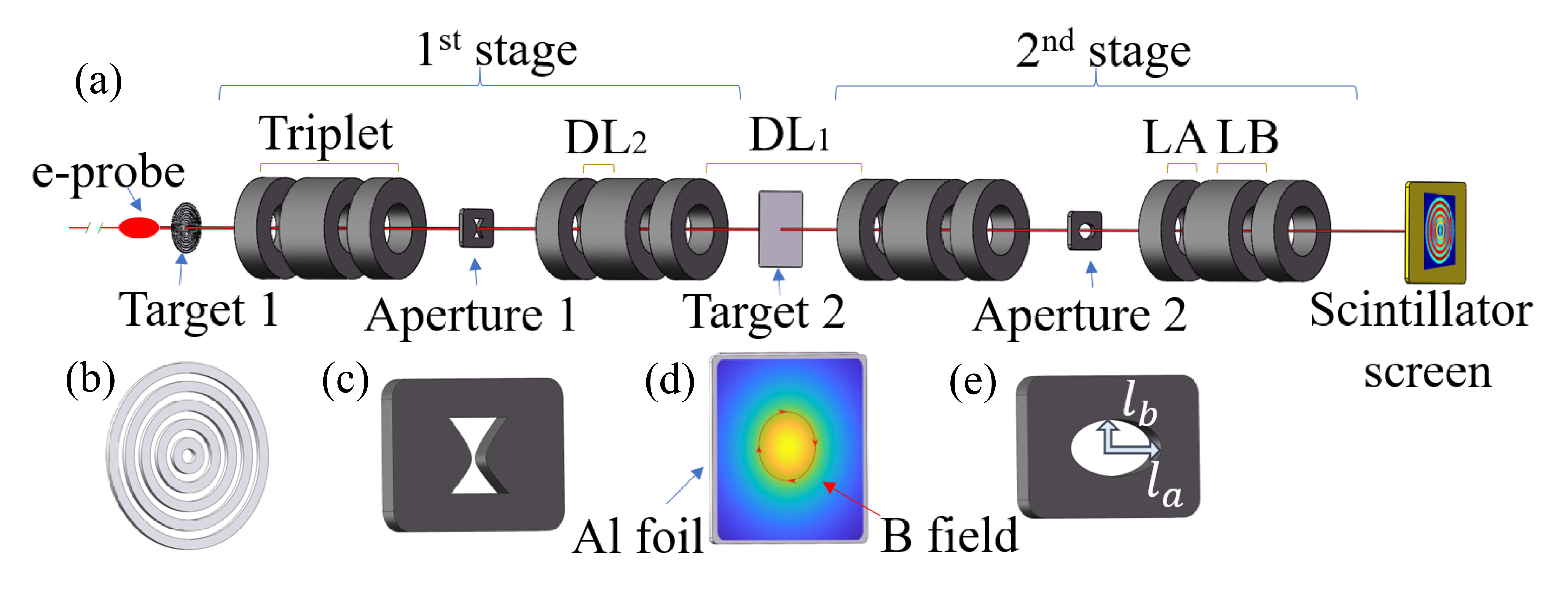}%
	\caption{\label{2}The simulation experiment setup. (a) The entire diagnostic system. (b) Scattering target 1, to create distinct areas as $\varphi'_{\bm{\theta}}$MB (filled with $10\mu m$ aluminum ) and $\rho'_{\bm{\theta}}$MB (empty area). (c) Aperture 1 at Fourier plane 1, to destroy the angle symmetry of the electrons passing through $\varphi'_{\bm{\theta}}$MB. (d) Diagnosed target consists of aluminum foil ($5\mu m$), which makes the beam has specific angle distribution, and circularly distributed magnetic field. (e) Aperture 2 is used to make angle selection at Fourier plane 2. }
\end{figure*}
Integrating the scattering, deflection and angle selection, the transmittance of electrons at the image plane is given by:
\begin{equation}
	\begin{aligned}	
		Tr({\bm{\xi}},{\bm{\theta}})		
		&=\iint\limits_{(x_F,y_F)\in{S}}g(x_F,y_F)dx_Fdy_F\\	
		&=\iint\limits_{(x'_0,y'_0)\in{S'}}j_{{\bm{\xi}}+{\bm{\theta}}}(x'_0,y'_0)dx'_0dy'_0\\	\label{Eq6}
		&=\iint\limits_{(\rho',\varphi')\in{S'}}f_{{\bm{\xi}}+{\bm{\theta}}}(\rho',\varphi')\rho'd\rho'd\varphi'
	\end{aligned}	
\end{equation}
where $ S $ and $ S' $ are the aperture area and its corresponding angle range. As previously stated, $f_{{\bm{\xi}}+{\bm{\theta}}}(\rho',\varphi')$ is identical to $f_{\bm{\xi}}(\rho')$ but with a shift. On that account, when $f_{\bm{\xi}}(\rho')$ is decided, if the aperture is set as an ellipse placed on the center of the beamline with $ l_a $ and $ l_b $ as the semiaxis of x-axis and y-axis respectively, and $ l_a/l_b=R_{12F}/R_{34F} $, the transmittance will depend on the field strength (determinant factor of $\rho'_{\bm{\theta}}$) and Eq.\ref{Eq6} can be rewritten as:
\begin{equation}
	\label{eq3}
	\begin{aligned}	Tr({\bm{\xi}},{\bm{\theta}})&=\iint\limits_{({\rho'},{\varphi'})\in{S'}}f_{\bm{\xi}}(\rho'_{\bm{OP}}){\rho'}d{\rho'}d\varphi'\\
		&=\int_0^{\varepsilon}\int_{0}^{2\pi}f_{\bm{\xi}}(\rho'_{\bm{OP}}){\rho'}d\rho'd\varphi'\\
	\end{aligned}
\end{equation}
where $\rho'_{\bm{OP}}$ is the radius with the center of $f_{{\bm{\xi}}+{\bm{\theta}}}(\rho',\varphi')$, $O$, as the pole, and
\begin{equation}
	\begin{aligned}
		\rho'_{\bm{OP}}=\sqrt{(\rho'cos\varphi'-\rho'_{\bm{\theta}}{cos\varphi'_{\bm{\theta}}})^2+(\rho'sin\varphi'-\rho'_{\bm{\theta}}{sin\varphi'_{\bm{\theta}}})^2} \label{eq1}
	\end{aligned}
\end{equation}
\begin{equation}\nonumber
	\begin{aligned}
		\varepsilon=l_a/R_{12F}=l_b/R_{34F}\
	\end{aligned}
\end{equation}
Furthermore, because of the circular symmetry of $f_{\bm{\xi}}(\rho')$, Eq.\ref{eq1} can be simplified as (see appendix A):
\begin{equation}
	\label{Eq2}
	\begin{aligned}
		\rho'_{\bm{OP}}=\sqrt{(\rho'cos\varphi'-\rho'_{\bm{\theta}})^2+(\rho'sin\varphi')^2}
	\end{aligned}
\end{equation}
For now, the correspondence between transmittance and field strength has been illustrated. This paves the way to achieve continuous measurement of field strength in transverse direction. 

Simplification from Eq.\ref{eq1} to Eq.\ref{Eq2} is based on the circular symmetry of $f_{\bm{\xi}}(\rho')$. Once the symmetry is destroyed, $\varphi'_{\bm{\theta}}$ in Eq.\ref{eq1} will be ineliminable, which gives the clue to make field direction detection. Dual Transmittance Diagnostics (DTD) is inspired by this. The principle is as shown in Fig.\ref{1}.
As shown in Fig.\ref{1}.b, 
the diagnosed area is divided into two sub-areas.
For the pink area, the angle distribution of beams propagating through is circularly symmetrical  (as shown in Fig.\ref{1}.c), its field strength can be determined by the transmittance at the image plane. This area is refereed as $\rho'_{\bm{\theta}}$-Measuring-Band($\rho'_{\bm{\theta}}$MB). For the blue area, the circular symmetry is destroyed (as shown in Fig.\ref{1}.e) so the transmittance is related to the field direction. This area is refereed as $\varphi'_{\bm{\theta}}$-Measuring-Band($\varphi'_{\bm{\theta}}$MB). By fitting the field strength from $\rho'_\theta$MB, the field strength of $\varphi'_{\bm{\theta}}$MB can be fixed. Vice versa, the field direction of $\rho'_\theta$MB can be fixed by fitting the transmittance of $\varphi'_{\bm{\theta}}$MB. Take point P as an example, its field strength can be determined by the transmittance of $\rho'_{\bm{\theta}}$MB and its field direction can be fixed by fitting transmittance of $\varphi'_{\bm{\theta}}$MB along the black line.

Preliminary simulation has been done to verify the feasibility of this diagnostics. The schematic diagram of this simulation setup is shown in Fig.\ref{2}. 
The entire system includes two sets of identical HEELR unit.
The first one is to destroy the angle symmetry of probe beams to create the $\varphi'_{\bm{\theta}}$MB, and the downstream one is the imaging unit. The beam optics for $50MeV$ electron radiography is designed and optimized via a high-order beam transport code COSY Infinity 9.1\cite{MAKINO2006346}. The total length of the system is $ 3.2m\times 2 $. The parameters of the beamline are listed in table \ref{table1}.
\mold{As shown in Fig.\ref{2}.d, the diagnosed target consists of aluminum foil and magnetic field.}

Aperture 1 is set as sandglass shape to break the angle symmetry but leave enough electrons to make a image. The diagnosed target consists of aluminum foil and magnetic field. The thickness of $B$ field area is $ 1\mu m $. Since it's more convenient to analyze in polar coordinate, the $B$ field is set clockwise from the electron's view, and its strength is radially decreasing from $ 680T*\mu m $ to $ 170T*\mu m $ (see appendix B).
\begin{table}[H]
	\caption{\label{table1}Parameters of the beamline.}
\begin{ruledtabular}
\begin{tabular}{cc}
			Parameters&Value\\
			\colrule
			$ QA,QB $&$ 3.589 T/m, -3.263 T/m $\\
			$ LA,LB $&$ 10cm,20cm $\\
			$ DL_1,DL_2 $&$ 50cm,10cm $\\
			$ l_a,l_b $&$ 1.044mm,0.744mm $\\
			$ R_{11F},R_{12F} $ &$ 0.00,1.044mm/mrad $ \\
			$ R_{33F},R_{34F} $ &$ 0.00,0.744mm/mrad$ \\
			$ R_{11} , R_{12}$	 &$ -1.00, 0.00mm/mrad $\\
			$ R_{33}, R_{34} $	 &$ -1.00,0.00mm/mrad  $\\
\end{tabular}
\end{ruledtabular}
\end{table}
According to the transfer matrix factors in table.\ref{table1}, the shape of aperture 2 is designed as an ellipse as shown in Fig \ref{2}.e and the corresponding collection angle range is:
\begin{equation}
	\sqrt{{x_{0}^{'}}^2+{y_{0}^{'}}^2}\le{l_a/R_{12F}}={l_b/R_{34F}}=1mrad
\end{equation}
The EGS5\cite{osti_877459} code is selected to simulate the interactions of electron with target.
The beam dynamic simulation is carried out by the ASTRA code\cite{Astra}.

\begin{figure}[hbp]
	\includegraphics[width=3.2in]{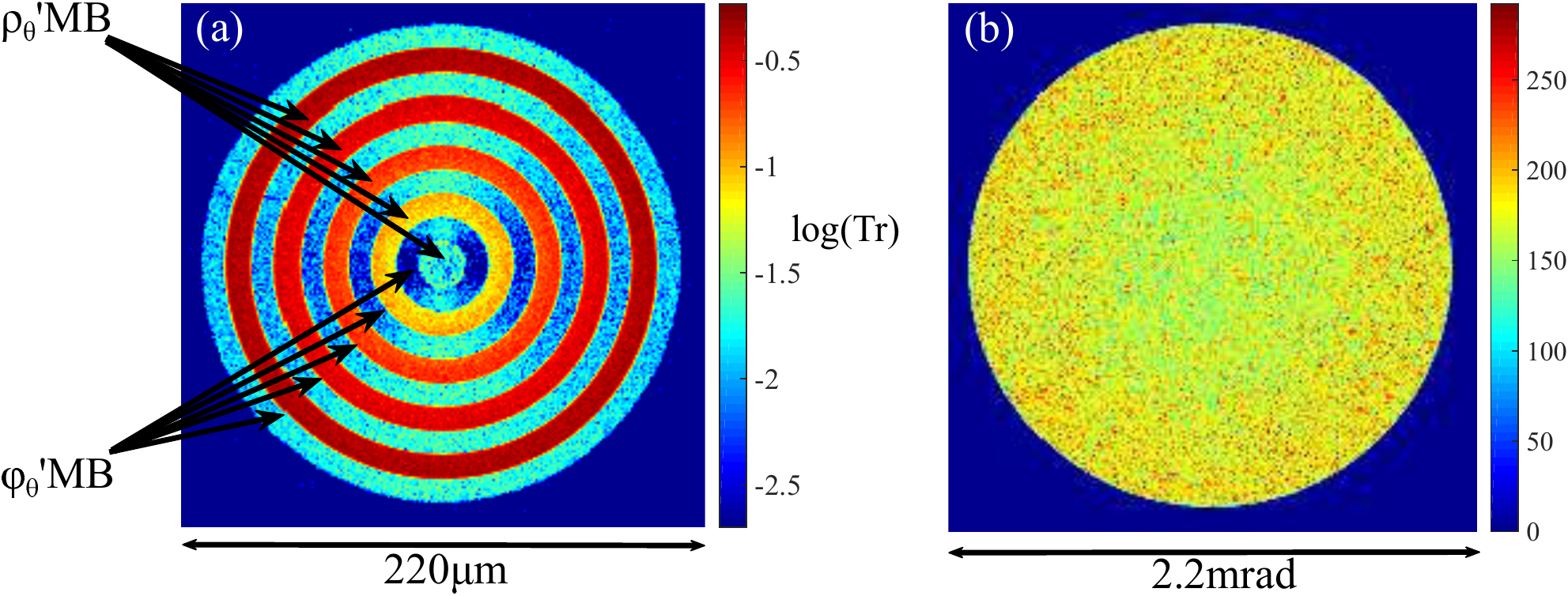}
	\caption{\label{fig3}(a)Electron beam profile at image plane. (b) $x'-y'$ distribution at image plane.}
\end{figure}

The image result is shown in Fig.\ref{fig3}.
In Fig.\ref{fig3}.a, two kinds of band are obvious. For the $\rho'_{\bm{\theta}}$MB area, the transmittance is static around the circle. However, in $\varphi'_{\bm{\theta}}$MB, the transmittance varies as the change of the  direction. Furthermore, as shown in Fig.\ref{fig3}.b, the electron angle distribution is confined in the specific range consistent with the designed value of aperture 2. To explain the principle more clearly, the two kinds of band in Fig.\ref{fig3}.(a) are discussed separately.
\begin{figure}[htp]
	\includegraphics[width=3.2in]{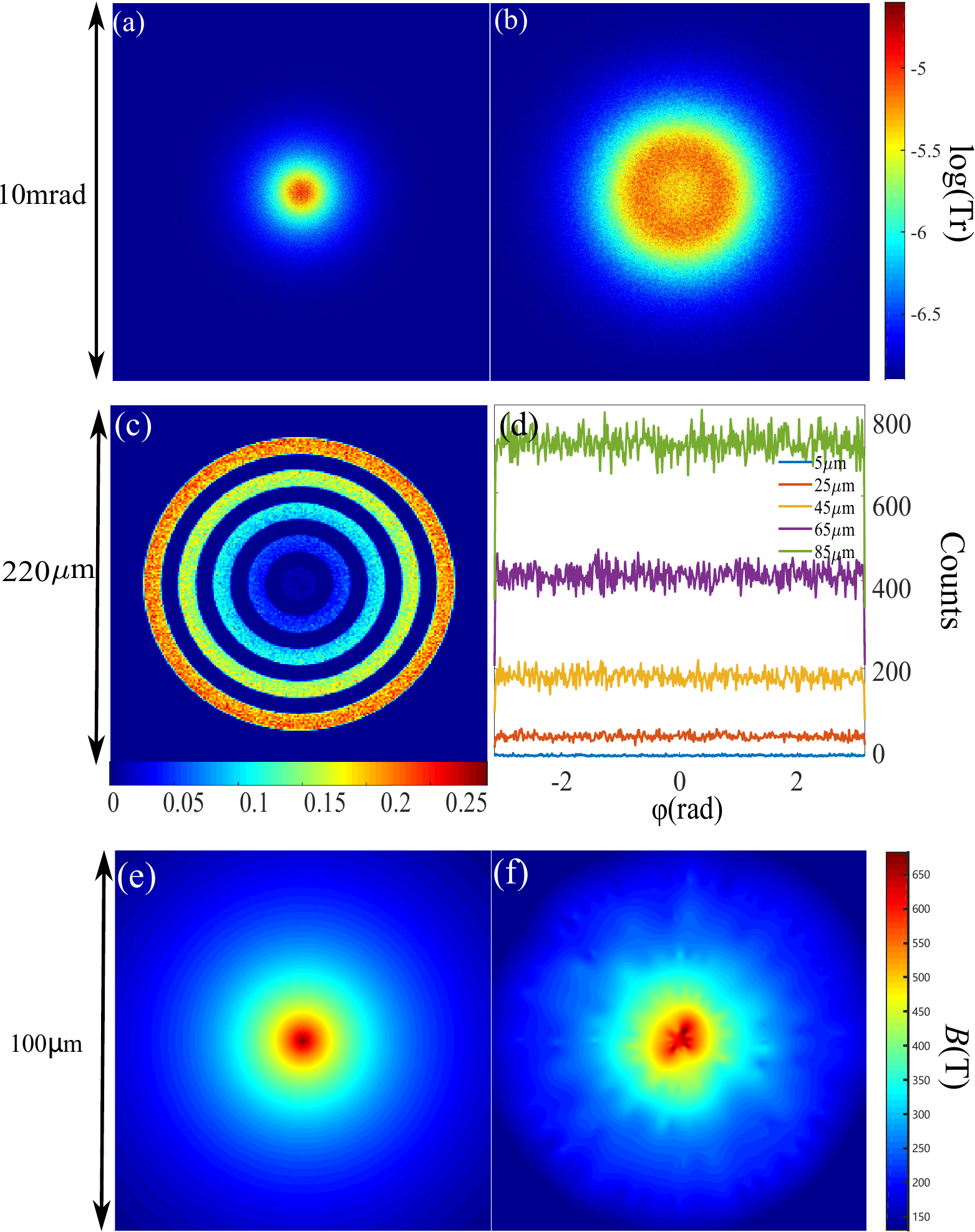}
	\caption{\label{4}The electron distribution of $\rho'_{\bm{\theta}}$MB at different position along the beamline. (a)$x'-y'$ distribution after aluminum foil of scattering target 2 but before the magnetic field. (b) $x'-y'$ distribution after the $B$ field. (c) Beam profile at the image plane. (d) Statistical result along angular direction of different bands in (c). (e) The pattern of the preset $B$ field. (f) Reconstruction result of $B$ field derived from (c).}
\end{figure}

For electrons penetrating $\rho'_{\bm{\theta}}$MB, since there is no scattering from target 1, aperture 1 will let them all through. Fig.\ref{4} shows the beam information at different position along the beamline. Contrast between Fig.\ref{4}.(a) and \ref{4}.(b) shows that the angle distribution of electrons is broadened by the $B$ field. Meanwhile, due to the circular symmetry of $B$ distribution, the deflection angle is circularly symmetrical as well. Consequently, the transmittance of any point at the image plane is supposed to be only related to the radius in polar coordinate. 

Fig.\ref{4}.(c) and (d) show that the transmittance increases as the distance to the center of the pattern grows, but remains consistent around each circle. This verified the prediction that the transmittance varies as the field strength grows but is independent from the field direction.
Fig.\ref{4}.(e) and (f) implies that the measured result agrees well with the preset value. When the field strength is the only quality we wanted, this result indicates that we can achieve point-by-point measurement actually.

\begin{figure}[htp]
	\includegraphics[width=3.2in]{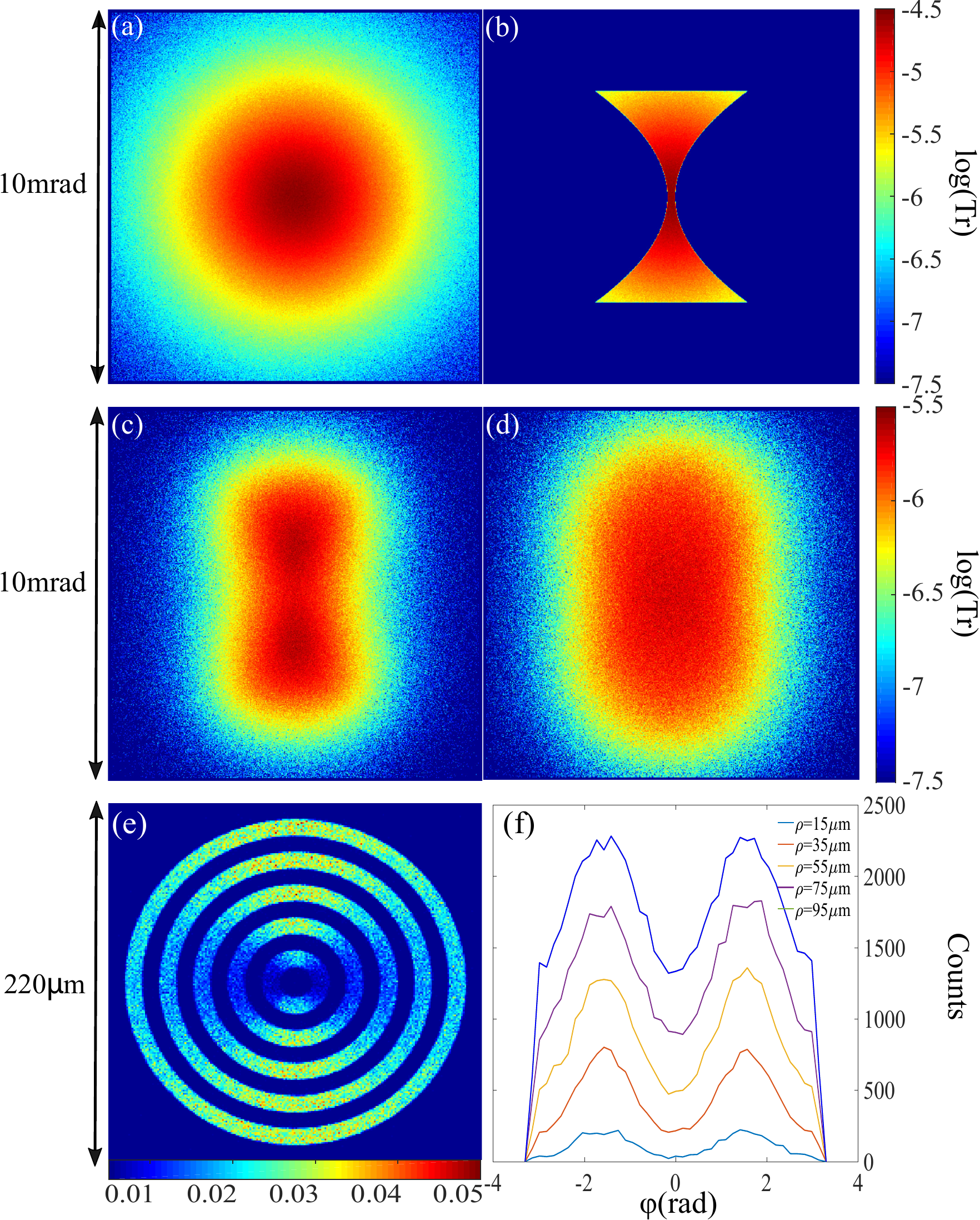}
	\caption{\label{5}The electron information of $\varphi'_{\bm{\theta}}$MB at different plane. (a) $x'-y'$ distribution after target 1. (b) $x'-y'$ distribution after aperture 1. (c) and (d) are the $x'-y'$ distributions before and behind $B$ field of target 2. (e) Beam profile at the image plane. (f) Statistical result along different bands in (e).}
\end{figure}

As for $\varphi'_{\bm{\theta}}$MB, as mentioned previously, the field strength can be obtained by fit and interpolation of $\rho'_{\bm{\theta}}$MB. On the other hand, unlike $\rho'_{\bm{\theta}}$MB, target 1 will scatter the electrons propagating through, so aperture 1 will work to break the circular symmetry of beam angle distribution by blocking electrons outside the hourglass shape. As shown in Fig.\ref{5}, the angle distribution changes from \ref{5}(a) to \ref{5}.(b) as electrons are filtered by aperture 1.
Fig.\ref{5}.(c) and \ref{5}.(d) correspond to Fig.\ref{4}.(a) and \ref{4}.(b) respectively. It's obvious that the angle distribution of electrons expanded as well after penetrating the $B$ field.
The statistical result of bands with different radius in \ref{5}.(e) is illustrated in \ref{5}.(f) which shows that the transmittance is different for the $B$ field with the same strength but different direction. In other words, when the field strength is known, the direction can be decided by the transmittance. This design can only obtain the field direction in one quadrant due to the rotational symmetry of aperture 1. Combining $\rho'_{\bm{\theta}}$MB and $\varphi'_{\bm{\theta}}$MB, the field direction and strength of any point at the image plane can be determined.

In summary, this letter proposed a diagnostic method which can get transverse electromagnetic field vector based on HEELR. The field strength can be measured conveniently and the field direction can be determined by DTD. The simulation of $ 50MeV $ electron beam as probe diagnosing the magnetic field ranging from $ 170T*\mu m $ to $\sim600T*\mu m $ validates the effectiveness of this method. Combined with the advantages of electron beams, this method is very suitable for diagnosis of ultrafast evolving electromagnetic field.\\
\begin{acknowledgments}
	This work is supported by the National Key R\&D Program of China under Grant (NO. SQ2019YFA040016). The authors thank Dr. P. Ch. Ai for advice on improving pictures in this article. \\
\end{acknowledgments}

%

\clearpage
\appendix

\section{APPENDIX A: DETAILED EXPLANATION OF EQUATION \ref{Eq2}}
\label{AppendixA}
\setcounter{equation}{0}
\renewcommand\theequation{A\arabic{equation}}
\setcounter{figure}{0}
\renewcommand{\thefigure}{A\arabic{figure}}
\begin{figure}[H]
	\includegraphics[width=3.2in]{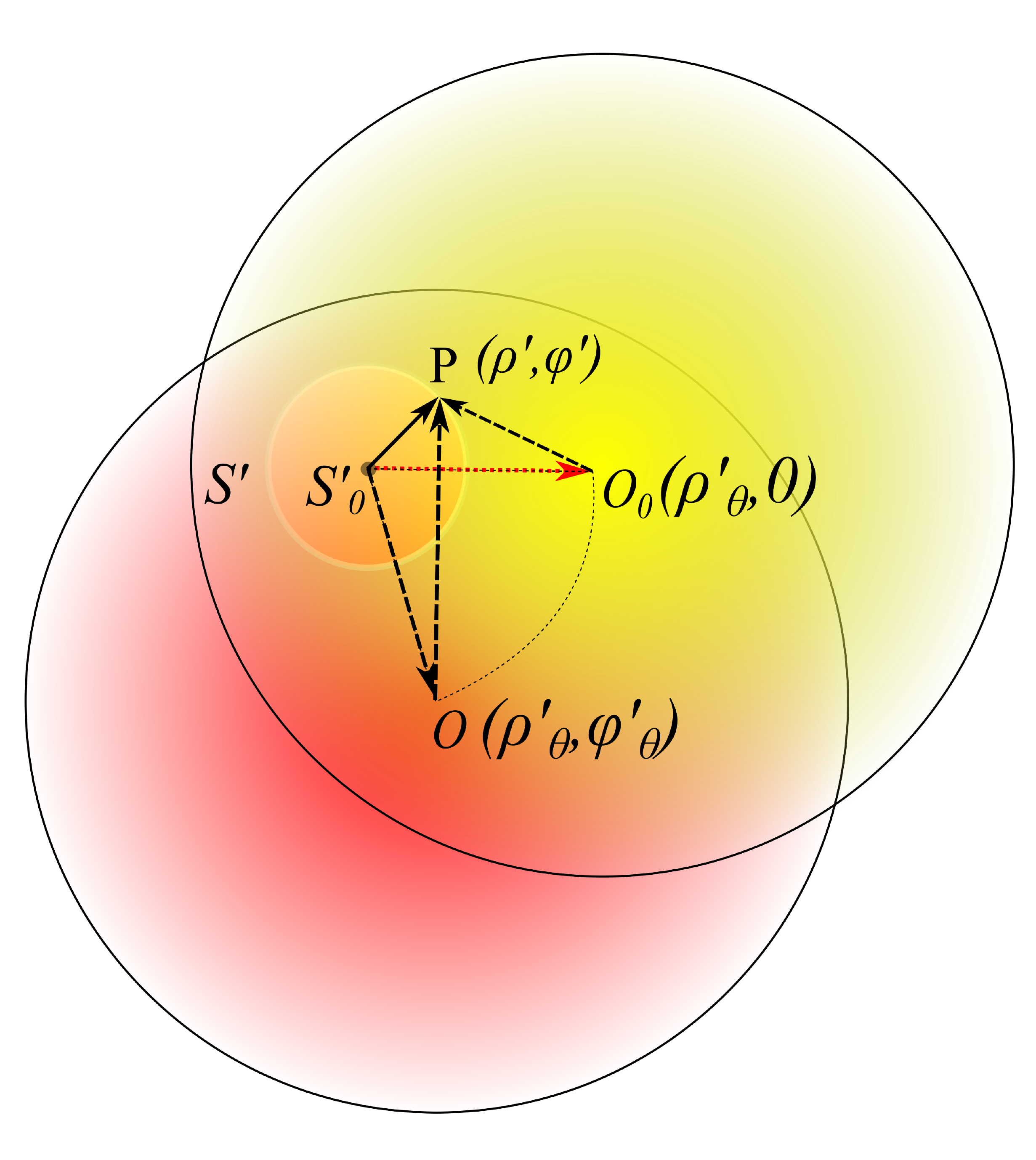}%
	\caption{\label{6}Cartoon concept of two  circular symmetrical  $f_{{\bm{\xi}}+{\bm{\theta}}}(\rho',\varphi')$ distribution with $O_0$ (yellow) and $O$ (red) as their center respectively. $S'_0$ is the center of $S'$, electrons out of $S'$ are blocked by aperture. The red dot line is the polar axis of polar coordinate system with $S'_0$ as the pole.}
\end{figure}
As shown in Fig.\ref{6}, there are two circular symmetrical  $f_{{\bm{\xi}}+{\bm{\theta}}}(\rho',\varphi')$ distribution with the same magnetic strength ($\rho'_{\bm{\theta}}$) but different direction ($\varphi'_{\bm{\theta}}$). In the polar coordinate system in Fig.\ref{6}, the coordinates of any point P in $S'$ is $ (\rho',\varphi') $, and in general situation the coordinates of $O$ is $ (\rho'_{\bm{\theta}},\varphi'_{{\bm{\theta}}}) $. The distance from point P to $O$ is:
\begin{equation}
	\begin{aligned}
		\rho'_{\bm{OP}}&=|\bm{OP}|\\ \label{eq11}
		&=|\bm{S'_0P}-\bm{S'_0O}|\\ 
		&=\sqrt{(\rho'cos\varphi'-\rho'_{\bm{\theta}}{cos\phi'_{{\bm{\theta}}}})^2+(\rho'sin\varphi'-\rho'_{\bm{\theta}}{sin\phi'_{{\bm{\theta}}}})^2}\\
	\end{aligned}
\end{equation}
Due to the circular symmetry, the integral result of Eq.\ref{Eq6} with $O$ centered distribution (red) is equivalent to the result with $O_0$ centered distribution (yellow). So Eq.\ref{eq11} can be simplified as:
\begin{equation}
	\rho'_{\bm{OP}}=\sqrt{(\rho'cos\varphi'-\rho'_{\bm{\theta}})^2+(\rho'sin\varphi')^2}
\end{equation}\
\section{APPENDIX B: DISTRIBUTION OF DIAGNOSED MAGNETIC FIELD}
\setcounter{figure}{0}
\renewcommand{\thefigure}{B\arabic{figure}}
\begin{figure}[H]
	\includegraphics[width=3.2in]{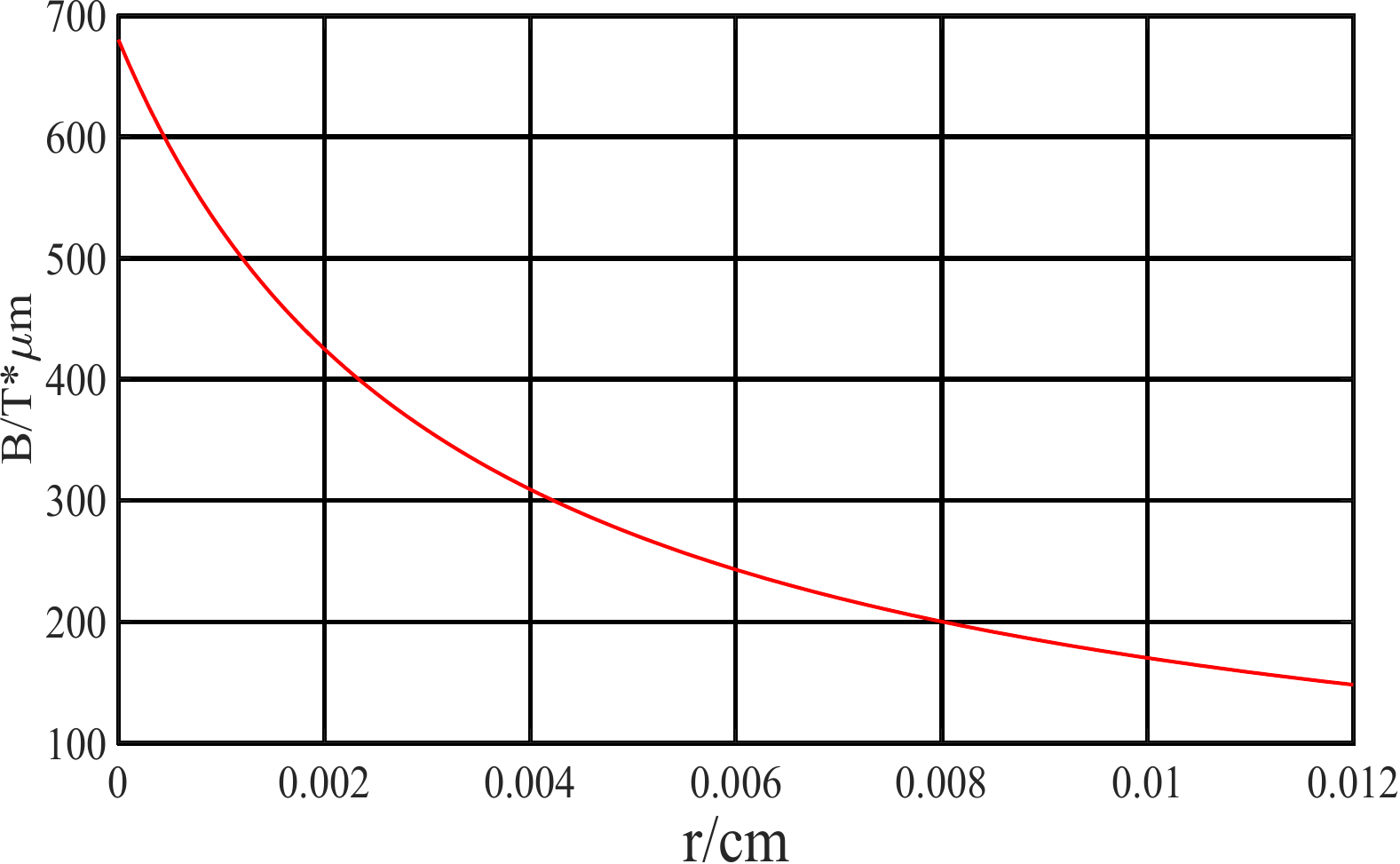}%
	\caption{\label{7} The $B$ field is set clockwise from the electron's view, and its strength is radially decreasing from $ 680T*\mu m $ to $ 170T*\mu m $.}
\end{figure}
The distribution of the diagnosed magnetic field in target 2 is shown above.

\begin{thebibliography}{29}%
	\makeatletter
	\providecommand \@ifxundefined [1]{%
		\@ifx{#1\undefined}
	}%
	\providecommand \@ifnum [1]{%
		\ifnum #1\expandafter \@firstoftwo
		\else \expandafter \@secondoftwo
		\fi
	}%
	\providecommand \@ifx [1]{%
		\ifx #1\expandafter \@firstoftwo
		\else \expandafter \@secondoftwo
		\fi
	}%
	\providecommand \natexlab [1]{#1}%
	\providecommand \enquote  [1]{``#1''}%
	\providecommand \bibnamefont  [1]{#1}%
	\providecommand \bibfnamefont [1]{#1}%
	\providecommand \citenamefont [1]{#1}%
	\providecommand \href@noop [0]{\@secondoftwo}%
	\providecommand \href [0]{\begingroup \@sanitize@url \@href}%
	\providecommand \@href[1]{\@@startlink{#1}\@@href}%
	\providecommand \@@href[1]{\endgroup#1\@@endlink}%
	\providecommand \@sanitize@url [0]{\catcode `\\12\catcode `\$12\catcode
		`\&12\catcode `\#12\catcode `\^12\catcode `\_12\catcode `\%12\relax}%
	\providecommand \@@startlink[1]{}%
	\providecommand \@@endlink[0]{}%
	\providecommand \url  [0]{\begingroup\@sanitize@url \@url }%
	\providecommand \@url [1]{\endgroup\@href {#1}{\urlprefix }}%
	\providecommand \urlprefix  [0]{URL }%
	\providecommand \Eprint [0]{\href }%
	\providecommand \doibase [0]{https://doi.org/}%
	\providecommand \selectlanguage [0]{\@gobble}%
	\providecommand \bibinfo  [0]{\@secondoftwo}%
	\providecommand \bibfield  [0]{\@secondoftwo}%
	\providecommand \translation [1]{[#1]}%
	\providecommand \BibitemOpen [0]{}%
	\providecommand \bibitemStop [0]{}%
	\providecommand \bibitemNoStop [0]{.\EOS\space}%
	\providecommand \EOS [0]{\spacefactor3000\relax}%
	\providecommand \BibitemShut  [1]{\csname bibitem#1\endcsname}%
	\let\auto@bib@innerbib\@empty
	\bibitem [{\citenamefont {Tajima}\ and\ \citenamefont
		{Dawson}(1979)}]{PhysRevLett.43.267}%
	\BibitemOpen
	\bibfield  {author} {\bibinfo {author} {\bibfnamefont {T.}~\bibnamefont
			{Tajima}}\ and\ \bibinfo {author} {\bibfnamefont {J.~M.}\ \bibnamefont
			{Dawson}},\ }\href {https://doi.org/10.1103/PhysRevLett.43.267} {\bibfield
		{journal} {\bibinfo  {journal} {Phys. Rev. Lett.}\ }\textbf {\bibinfo
			{volume} {43}},\ \bibinfo {pages} {267} (\bibinfo {year} {1979})}\BibitemShut
	{NoStop}%
	\bibitem [{\citenamefont {Haberberger}\ \emph {et~al.}(2012)\citenamefont
		{Haberberger}, \citenamefont {Tochitsky}, \citenamefont {Fiuza},
		\citenamefont {Gong}, \citenamefont {Fonseca}, \citenamefont {Silva},
		\citenamefont {Mori},\ and\ \citenamefont {Joshi}}]{article2}%
	\BibitemOpen
	\bibfield  {author} {\bibinfo {author} {\bibfnamefont {D.}~\bibnamefont
			{Haberberger}}, \bibinfo {author} {\bibfnamefont {S.}~\bibnamefont
			{Tochitsky}}, \bibinfo {author} {\bibfnamefont {F.}~\bibnamefont {Fiuza}},
		\bibinfo {author} {\bibfnamefont {C.}~\bibnamefont {Gong}}, \bibinfo {author}
		{\bibfnamefont {R.}~\bibnamefont {Fonseca}}, \bibinfo {author} {\bibfnamefont
			{L.}~\bibnamefont {Silva}}, \bibinfo {author} {\bibfnamefont
			{W.}~\bibnamefont {Mori}},\ and\ \bibinfo {author} {\bibfnamefont
			{C.}~\bibnamefont {Joshi}},\ }\href {https://doi.org/10.1038/nphys2130}
	{\bibfield  {journal} {\bibinfo  {journal} {Nature Physics}\ }\textbf
		{\bibinfo {volume} {8}},\ \bibinfo {pages} {95} (\bibinfo {year}
		{2012})}\BibitemShut {NoStop}%
	\bibitem [{\citenamefont {Haines}(1997)}]{PhysRevLett.78.254}%
	\BibitemOpen
	\bibfield  {author} {\bibinfo {author} {\bibfnamefont {M.~G.}\ \bibnamefont
			{Haines}},\ }\href {https://doi.org/10.1103/PhysRevLett.78.254} {\bibfield
		{journal} {\bibinfo  {journal} {Phys. Rev. Lett.}\ }\textbf {\bibinfo
			{volume} {78}},\ \bibinfo {pages} {254} (\bibinfo {year} {1997})}\BibitemShut
	{NoStop}%
	\bibitem [{\citenamefont {Jr}(1962)}]{article4}%
	\BibitemOpen
	\bibfield  {author} {\bibinfo {author} {\bibfnamefont {L.}~\bibnamefont
			{Jr}},\ }\href@noop {} {\emph {\bibinfo {title} {Physics of Fully Ionized
				Gases, 2nd Edition}}}\ (\bibinfo  {publisher} {Wiley-Interscience},\ \bibinfo
	{year} {1962})\BibitemShut {NoStop}%
	\bibitem [{\citenamefont {{Braginskii}}(1965)}]{book1}%
	\BibitemOpen
	\bibfield  {author} {\bibinfo {author} {\bibfnamefont {S.~I.}\ \bibnamefont
			{{Braginskii}}},\ }\href@noop {} {\bibfield  {journal} {\bibinfo  {journal}
			{Reviews of Plasma Physics}\ }\textbf {\bibinfo {volume} {1}},\ \bibinfo
		{pages} {205} (\bibinfo {year} {1965})}\BibitemShut {NoStop}%
	\bibitem [{\citenamefont {Walsh}\ \emph {et~al.}(2017)\citenamefont {Walsh},
		\citenamefont {Chittenden}, \citenamefont {McGlinchey}, \citenamefont
		{Niasse},\ and\ \citenamefont {Appelbe}}]{PhysRevLett.118.155001}%
	\BibitemOpen
	\bibfield  {author} {\bibinfo {author} {\bibfnamefont {C.~A.}\ \bibnamefont
			{Walsh}}, \bibinfo {author} {\bibfnamefont {J.~P.}\ \bibnamefont
			{Chittenden}}, \bibinfo {author} {\bibfnamefont {K.}~\bibnamefont
			{McGlinchey}}, \bibinfo {author} {\bibfnamefont {N.~P.~L.}\ \bibnamefont
			{Niasse}},\ and\ \bibinfo {author} {\bibfnamefont {B.~D.}\ \bibnamefont
			{Appelbe}},\ }\href {https://doi.org/10.1103/PhysRevLett.118.155001}
	{\bibfield  {journal} {\bibinfo  {journal} {Phys. Rev. Lett.}\ }\textbf
		{\bibinfo {volume} {118}},\ \bibinfo {pages} {155001} (\bibinfo {year}
		{2017})}\BibitemShut {NoStop}%
	\bibitem [{\citenamefont {Gotchev}\ \emph {et~al.}(2009)\citenamefont
		{Gotchev}, \citenamefont {Chang}, \citenamefont {Knauer}, \citenamefont
		{Meyerhofer}, \citenamefont {Polomarov}, \citenamefont {Frenje},
		\citenamefont {Li}, \citenamefont {Manuel}, \citenamefont {Petrasso},
		\citenamefont {Rygg} \emph {et~al.}}]{PhysRevLett.103.215004}%
	\BibitemOpen
	\bibfield  {author} {\bibinfo {author} {\bibfnamefont {O.~V.}\ \bibnamefont
			{Gotchev}}, \bibinfo {author} {\bibfnamefont {P.~Y.}\ \bibnamefont {Chang}},
		\bibinfo {author} {\bibfnamefont {J.~P.}\ \bibnamefont {Knauer}}, \bibinfo
		{author} {\bibfnamefont {D.~D.}\ \bibnamefont {Meyerhofer}}, \bibinfo
		{author} {\bibfnamefont {O.}~\bibnamefont {Polomarov}}, \bibinfo {author}
		{\bibfnamefont {J.}~\bibnamefont {Frenje}}, \bibinfo {author} {\bibfnamefont
			{C.~K.}\ \bibnamefont {Li}}, \bibinfo {author} {\bibfnamefont {M.~J.-E.}\
			\bibnamefont {Manuel}}, \bibinfo {author} {\bibfnamefont {R.~D.}\
			\bibnamefont {Petrasso}}, \bibinfo {author} {\bibfnamefont {J.~R.}\
			\bibnamefont {Rygg}}, \emph {et~al.},\ }\href
	{https://doi.org/10.1103/PhysRevLett.103.215004} {\bibfield  {journal}
		{\bibinfo  {journal} {Phys. Rev. Lett.}\ }\textbf {\bibinfo {volume} {103}},\
		\bibinfo {pages} {215004} (\bibinfo {year} {2009})}\BibitemShut {NoStop}%
	\bibitem [{\citenamefont {{Lindemuth}}\ \emph {et~al.}(1997)\citenamefont
		{{Lindemuth}}, \citenamefont {{Ekdahl}}, \citenamefont {{Fowler}},
		\citenamefont {{Reinovsky}}, \citenamefont {{Younger}}, \citenamefont
		{{Chernyshev}}, \citenamefont {{Mokhov}},\ and\ \citenamefont
		{{Pavlovskii}}}]{650905}%
	\BibitemOpen
	\bibfield  {author} {\bibinfo {author} {\bibfnamefont {L.~R.}\ \bibnamefont
			{{Lindemuth}}}, \bibinfo {author} {\bibfnamefont {C.~A.}\ \bibnamefont
			{{Ekdahl}}}, \bibinfo {author} {\bibfnamefont {C.~M.}\ \bibnamefont
			{{Fowler}}}, \bibinfo {author} {\bibfnamefont {R.~E.}\ \bibnamefont
			{{Reinovsky}}}, \bibinfo {author} {\bibfnamefont {S.~M.}\ \bibnamefont
			{{Younger}}}, \bibinfo {author} {\bibfnamefont {V.~K.}\ \bibnamefont
			{{Chernyshev}}}, \bibinfo {author} {\bibfnamefont {V.~N.}\ \bibnamefont
			{{Mokhov}}},\ and\ \bibinfo {author} {\bibfnamefont {A.~I.}\ \bibnamefont
			{{Pavlovskii}}},\ }\href {https://doi.org/10.1109/27.650905} {\bibfield
		{journal} {\bibinfo  {journal} {IEEE Transactions on Plasma Science}\
		}\textbf {\bibinfo {volume} {25}},\ \bibinfo {pages} {1357} (\bibinfo {year}
		{1997})}\BibitemShut {NoStop}%
	\bibitem [{\citenamefont {Nilson}\ \emph {et~al.}(2006)\citenamefont {Nilson},
		\citenamefont {Willingale}, \citenamefont {Kaluza}, \citenamefont
		{Kamperidis}, \citenamefont {Minardi}, \citenamefont {Wei}, \citenamefont
		{Fernandes}, \citenamefont {Notley}, \citenamefont {Bandyopadhyay},
		\citenamefont {Sherlock} \emph {et~al.}}]{PhysRevLett.97.255001}%
	\BibitemOpen
	\bibfield  {author} {\bibinfo {author} {\bibfnamefont {P.~M.}\ \bibnamefont
			{Nilson}}, \bibinfo {author} {\bibfnamefont {L.}~\bibnamefont {Willingale}},
		\bibinfo {author} {\bibfnamefont {M.~C.}\ \bibnamefont {Kaluza}}, \bibinfo
		{author} {\bibfnamefont {C.}~\bibnamefont {Kamperidis}}, \bibinfo {author}
		{\bibfnamefont {S.}~\bibnamefont {Minardi}}, \bibinfo {author} {\bibfnamefont
			{M.~S.}\ \bibnamefont {Wei}}, \bibinfo {author} {\bibfnamefont
			{P.}~\bibnamefont {Fernandes}}, \bibinfo {author} {\bibfnamefont
			{M.}~\bibnamefont {Notley}}, \bibinfo {author} {\bibfnamefont
			{S.}~\bibnamefont {Bandyopadhyay}}, \bibinfo {author} {\bibfnamefont
			{M.}~\bibnamefont {Sherlock}}, \emph {et~al.},\ }\href
	{https://doi.org/10.1103/PhysRevLett.97.255001} {\bibfield  {journal}
		{\bibinfo  {journal} {Phys. Rev. Lett.}\ }\textbf {\bibinfo {volume} {97}},\
		\bibinfo {pages} {255001} (\bibinfo {year} {2006})}\BibitemShut {NoStop}%
	\bibitem [{\citenamefont {Fox}\ \emph {et~al.}(2012)\citenamefont {Fox},
		\citenamefont {Bhattacharjee},\ and\ \citenamefont
		{Germaschewski}}]{doi:10.1063/1.3694119}%
	\BibitemOpen
	\bibfield  {author} {\bibinfo {author} {\bibfnamefont {W.}~\bibnamefont
			{Fox}}, \bibinfo {author} {\bibfnamefont {A.}~\bibnamefont {Bhattacharjee}},\
		and\ \bibinfo {author} {\bibfnamefont {K.}~\bibnamefont {Germaschewski}},\
	}\href {https://doi.org/10.1063/1.3694119} {\bibfield  {journal} {\bibinfo
			{journal} {Physics of Plasmas}\ }\textbf {\bibinfo {volume} {19}},\ \bibinfo
		{pages} {056309} (\bibinfo {year} {2012})}\BibitemShut {NoStop}%
	\bibitem [{\citenamefont {Li}\ \emph {et~al.}(2006)\citenamefont {Li},
		\citenamefont {S\'eguin}, \citenamefont {Frenje}, \citenamefont {Rygg},
		\citenamefont {Petrasso}, \citenamefont {Town}, \citenamefont {Amendt},
		\citenamefont {Hatchett}, \citenamefont {Landen}, \citenamefont {Mackinnon}
		\emph {et~al.}}]{PhysRevLett.97.135003}%
	\BibitemOpen
	\bibfield  {author} {\bibinfo {author} {\bibfnamefont {C.~K.}\ \bibnamefont
			{Li}}, \bibinfo {author} {\bibfnamefont {F.~H.}\ \bibnamefont {S\'eguin}},
		\bibinfo {author} {\bibfnamefont {J.~A.}\ \bibnamefont {Frenje}}, \bibinfo
		{author} {\bibfnamefont {J.~R.}\ \bibnamefont {Rygg}}, \bibinfo {author}
		{\bibfnamefont {R.~D.}\ \bibnamefont {Petrasso}}, \bibinfo {author}
		{\bibfnamefont {R.~P.~J.}\ \bibnamefont {Town}}, \bibinfo {author}
		{\bibfnamefont {P.~A.}\ \bibnamefont {Amendt}}, \bibinfo {author}
		{\bibfnamefont {S.~P.}\ \bibnamefont {Hatchett}}, \bibinfo {author}
		{\bibfnamefont {O.~L.}\ \bibnamefont {Landen}}, \bibinfo {author}
		{\bibfnamefont {A.~J.}\ \bibnamefont {Mackinnon}}, \emph {et~al.},\ }\href
	{https://doi.org/10.1103/PhysRevLett.97.135003} {\bibfield  {journal}
		{\bibinfo  {journal} {Phys. Rev. Lett.}\ }\textbf {\bibinfo {volume} {97}},\
		\bibinfo {pages} {135003} (\bibinfo {year} {2006})}\BibitemShut {NoStop}%
	\bibitem [{\citenamefont {Li}\ \emph {et~al.}(2008)\citenamefont {Li},
		\citenamefont {S\'eguin}, \citenamefont {Rygg}, \citenamefont {Frenje},
		\citenamefont {Manuel}, \citenamefont {Petrasso}, \citenamefont {Betti},
		\citenamefont {Delettrez}, \citenamefont {Knauer}, \citenamefont {Marshall}
		\emph {et~al.}}]{PhysRevLett.100.225001}%
	\BibitemOpen
	\bibfield  {author} {\bibinfo {author} {\bibfnamefont {C.~K.}\ \bibnamefont
			{Li}}, \bibinfo {author} {\bibfnamefont {F.~H.}\ \bibnamefont {S\'eguin}},
		\bibinfo {author} {\bibfnamefont {J.~R.}\ \bibnamefont {Rygg}}, \bibinfo
		{author} {\bibfnamefont {J.~A.}\ \bibnamefont {Frenje}}, \bibinfo {author}
		{\bibfnamefont {M.}~\bibnamefont {Manuel}}, \bibinfo {author} {\bibfnamefont
			{R.~D.}\ \bibnamefont {Petrasso}}, \bibinfo {author} {\bibfnamefont
			{R.}~\bibnamefont {Betti}}, \bibinfo {author} {\bibfnamefont
			{J.}~\bibnamefont {Delettrez}}, \bibinfo {author} {\bibfnamefont {J.~P.}\
			\bibnamefont {Knauer}}, \bibinfo {author} {\bibfnamefont {F.}~\bibnamefont
			{Marshall}}, \emph {et~al.},\ }\href
	{https://doi.org/10.1103/PhysRevLett.100.225001} {\bibfield  {journal}
		{\bibinfo  {journal} {Phys. Rev. Lett.}\ }\textbf {\bibinfo {volume} {100}},\
		\bibinfo {pages} {225001} (\bibinfo {year} {2008})}\BibitemShut {NoStop}%
	\bibitem [{\citenamefont {Rygg}\ \emph {et~al.}(2008)\citenamefont {Rygg},
		\citenamefont {S{\'e}guin}, \citenamefont {Li}, \citenamefont {Frenje},
		\citenamefont {Manuel}, \citenamefont {Petrasso}, \citenamefont {Betti},
		\citenamefont {Delettrez}, \citenamefont {Gotchev}, \citenamefont {Knauer}
		\emph {et~al.}}]{Rygg1223}%
	\BibitemOpen
	\bibfield  {author} {\bibinfo {author} {\bibfnamefont {J.~R.}\ \bibnamefont
			{Rygg}}, \bibinfo {author} {\bibfnamefont {F.~H.}\ \bibnamefont
			{S{\'e}guin}}, \bibinfo {author} {\bibfnamefont {C.~K.}\ \bibnamefont {Li}},
		\bibinfo {author} {\bibfnamefont {J.~A.}\ \bibnamefont {Frenje}}, \bibinfo
		{author} {\bibfnamefont {M.~J.-E.}\ \bibnamefont {Manuel}}, \bibinfo {author}
		{\bibfnamefont {R.~D.}\ \bibnamefont {Petrasso}}, \bibinfo {author}
		{\bibfnamefont {R.}~\bibnamefont {Betti}}, \bibinfo {author} {\bibfnamefont
			{J.~A.}\ \bibnamefont {Delettrez}}, \bibinfo {author} {\bibfnamefont {O.~V.}\
			\bibnamefont {Gotchev}}, \bibinfo {author} {\bibfnamefont {J.~P.}\
			\bibnamefont {Knauer}}, \emph {et~al.},\ }\href
	{https://doi.org/10.1126/science.1152640} {\bibfield  {journal} {\bibinfo
			{journal} {Science}\ }\textbf {\bibinfo {volume} {319}},\ \bibinfo {pages}
		{1223} (\bibinfo {year} {2008})}\BibitemShut {NoStop}%
	\bibitem [{\citenamefont {Schumaker}\ \emph {et~al.}(2013)\citenamefont
		{Schumaker}, \citenamefont {Nakanii}, \citenamefont {McGuffey}, \citenamefont
		{Zulick}, \citenamefont {Chyvkov}, \citenamefont {Dollar}, \citenamefont
		{Habara}, \citenamefont {Kalintchenko}, \citenamefont {Maksimchuk},
		\citenamefont {Tanaka} \emph {et~al.}}]{PhysRevLett.110.015003}%
	\BibitemOpen
	\bibfield  {author} {\bibinfo {author} {\bibfnamefont {W.}~\bibnamefont
			{Schumaker}}, \bibinfo {author} {\bibfnamefont {N.}~\bibnamefont {Nakanii}},
		\bibinfo {author} {\bibfnamefont {C.}~\bibnamefont {McGuffey}}, \bibinfo
		{author} {\bibfnamefont {C.}~\bibnamefont {Zulick}}, \bibinfo {author}
		{\bibfnamefont {V.}~\bibnamefont {Chyvkov}}, \bibinfo {author} {\bibfnamefont
			{F.}~\bibnamefont {Dollar}}, \bibinfo {author} {\bibfnamefont
			{H.}~\bibnamefont {Habara}}, \bibinfo {author} {\bibfnamefont
			{G.}~\bibnamefont {Kalintchenko}}, \bibinfo {author} {\bibfnamefont
			{A.}~\bibnamefont {Maksimchuk}}, \bibinfo {author} {\bibfnamefont {K.~A.}\
			\bibnamefont {Tanaka}}, \emph {et~al.},\ }\href
	{https://doi.org/10.1103/PhysRevLett.110.015003} {\bibfield  {journal}
		{\bibinfo  {journal} {Phys. Rev. Lett.}\ }\textbf {\bibinfo {volume} {110}},\
		\bibinfo {pages} {015003} (\bibinfo {year} {2013})}\BibitemShut {NoStop}%
	\bibitem [{\citenamefont {Mottershead}\ and\ \citenamefont
		{Zumbro}(1997)}]{750705}%
	\BibitemOpen
	\bibfield  {author} {\bibinfo {author} {\bibfnamefont {C.}~\bibnamefont
			{Mottershead}}\ and\ \bibinfo {author} {\bibfnamefont {J.}~\bibnamefont
			{Zumbro}},\ }in\ \href {https://doi.org/10.1109/PAC.1997.750705} {\emph
		{\bibinfo {booktitle} {Proceedings of the 1997 Particle Accelerator
				Conference}}},\ Vol.~\bibinfo {volume} {2}\ (\bibinfo {year} {1997})\ pp.\
	\bibinfo {pages} {1397--1399}\BibitemShut {NoStop}%
	\bibitem [{\citenamefont {Sjue}\ \emph {et~al.}(2016)\citenamefont {Sjue},
		\citenamefont {Mariam}, \citenamefont {Merrill}, \citenamefont {Morris},\
		and\ \citenamefont {Saunders}}]{doi:10.1063/1.4939822}%
	\BibitemOpen
	\bibfield  {author} {\bibinfo {author} {\bibfnamefont {S.~K.~L.}\
			\bibnamefont {Sjue}}, \bibinfo {author} {\bibfnamefont {F.~G.}\ \bibnamefont
			{Mariam}}, \bibinfo {author} {\bibfnamefont {F.~E.}\ \bibnamefont {Merrill}},
		\bibinfo {author} {\bibfnamefont {C.~L.}\ \bibnamefont {Morris}},\ and\
		\bibinfo {author} {\bibfnamefont {A.}~\bibnamefont {Saunders}},\ }\href
	{https://doi.org/10.1063/1.4939822} {\bibfield  {journal} {\bibinfo
			{journal} {Review of Scientific Instruments}\ }\textbf {\bibinfo {volume}
			{87}},\ \bibinfo {pages} {015110} (\bibinfo {year} {2016})}\BibitemShut
	{NoStop}%
	\bibitem [{\citenamefont {Merrill}\ \emph {et~al.}(2017)\citenamefont
		{Merrill}, \citenamefont {Fabritius}, \citenamefont {Mariam}, \citenamefont
		{Poulson}, \citenamefont {Simpson}, \citenamefont {Walstrom},\ and\
		\citenamefont {Wilde}}]{doi:10.1063/1.4971560}%
	\BibitemOpen
	\bibfield  {author} {\bibinfo {author} {\bibfnamefont {F.~E.}\ \bibnamefont
			{Merrill}}, \bibinfo {author} {\bibfnamefont {J.}~\bibnamefont {Fabritius}},
		\bibinfo {author} {\bibfnamefont {F.~G.}\ \bibnamefont {Mariam}}, \bibinfo
		{author} {\bibfnamefont {D.}~\bibnamefont {Poulson}}, \bibinfo {author}
		{\bibfnamefont {R.}~\bibnamefont {Simpson}}, \bibinfo {author} {\bibfnamefont
			{P.}~\bibnamefont {Walstrom}},\ and\ \bibinfo {author} {\bibfnamefont
			{C.}~\bibnamefont {Wilde}},\ }\href {https://doi.org/10.1063/1.4971560}
	{\bibfield  {journal} {\bibinfo  {journal} {AIP Conference Proceedings}\
		}\textbf {\bibinfo {volume} {1793}},\ \bibinfo {pages} {060004} (\bibinfo
		{year} {2017})}\BibitemShut {NoStop}%
	\bibitem [{\citenamefont {Zhou}\ \emph
		{et~al.}(2019{\natexlab{a}})\citenamefont {Zhou}, \citenamefont {Fang},
		\citenamefont {Chen}, \citenamefont {Wu}, \citenamefont {Du}, \citenamefont
		{Yan}, \citenamefont {Tang},\ and\ \citenamefont
		{Huang}}]{PhysRevApplied.11.034068}%
	\BibitemOpen
	\bibfield  {author} {\bibinfo {author} {\bibfnamefont {Z.}~\bibnamefont
			{Zhou}}, \bibinfo {author} {\bibfnamefont {Y.}~\bibnamefont {Fang}}, \bibinfo
		{author} {\bibfnamefont {H.}~\bibnamefont {Chen}}, \bibinfo {author}
		{\bibfnamefont {Y.}~\bibnamefont {Wu}}, \bibinfo {author} {\bibfnamefont
			{Y.}~\bibnamefont {Du}}, \bibinfo {author} {\bibfnamefont {L.}~\bibnamefont
			{Yan}}, \bibinfo {author} {\bibfnamefont {C.}~\bibnamefont {Tang}},\ and\
		\bibinfo {author} {\bibfnamefont {W.}~\bibnamefont {Huang}},\ }\href
	{https://doi.org/10.1103/PhysRevApplied.11.034068} {\bibfield  {journal}
		{\bibinfo  {journal} {Phys. Rev. Applied}\ }\textbf {\bibinfo {volume}
			{11}},\ \bibinfo {pages} {034068} (\bibinfo {year}
		{2019}{\natexlab{a}})}\BibitemShut {NoStop}%
	\bibitem [{\citenamefont {Zhou}\ \emph
		{et~al.}(2019{\natexlab{b}})\citenamefont {Zhou}, \citenamefont {Fang},
		\citenamefont {Chen}, \citenamefont {Wu}, \citenamefont {Du}, \citenamefont
		{Zhang}, \citenamefont {Zhao}, \citenamefont {Li}, \citenamefont {Tang},\
		and\ \citenamefont {Huang}}]{doi:10.1063/1.5109855}%
	\BibitemOpen
	\bibfield  {author} {\bibinfo {author} {\bibfnamefont {Z.}~\bibnamefont
			{Zhou}}, \bibinfo {author} {\bibfnamefont {Y.}~\bibnamefont {Fang}}, \bibinfo
		{author} {\bibfnamefont {H.}~\bibnamefont {Chen}}, \bibinfo {author}
		{\bibfnamefont {Y.}~\bibnamefont {Wu}}, \bibinfo {author} {\bibfnamefont
			{Y.}~\bibnamefont {Du}}, \bibinfo {author} {\bibfnamefont {Z.}~\bibnamefont
			{Zhang}}, \bibinfo {author} {\bibfnamefont {Y.}~\bibnamefont {Zhao}},
		\bibinfo {author} {\bibfnamefont {M.}~\bibnamefont {Li}}, \bibinfo {author}
		{\bibfnamefont {C.}~\bibnamefont {Tang}},\ and\ \bibinfo {author}
		{\bibfnamefont {W.}~\bibnamefont {Huang}},\ }\href
	{https://doi.org/10.1063/1.5109855} {\bibfield  {journal} {\bibinfo
			{journal} {Matter and Radiation at Extremes}\ }\textbf {\bibinfo {volume}
			{4}},\ \bibinfo {pages} {065402} (\bibinfo {year}
		{2019}{\natexlab{b}})}\BibitemShut {NoStop}%
	\bibitem [{\citenamefont {Merrill}\ \emph {et~al.}(2018)\citenamefont
		{Merrill}, \citenamefont {Goett}, \citenamefont {Gibbs}, \citenamefont
		{Imhoff}, \citenamefont {Mariam}, \citenamefont {Morris}, \citenamefont
		{Neukirch}, \citenamefont {Perry}, \citenamefont {Poulson}, \citenamefont
		{Simpson} \emph {et~al.}}]{doi:10.1063/1.5011198}%
	\BibitemOpen
	\bibfield  {author} {\bibinfo {author} {\bibfnamefont {F.~E.}\ \bibnamefont
			{Merrill}}, \bibinfo {author} {\bibfnamefont {J.}~\bibnamefont {Goett}},
		\bibinfo {author} {\bibfnamefont {J.~W.}\ \bibnamefont {Gibbs}}, \bibinfo
		{author} {\bibfnamefont {S.~D.}\ \bibnamefont {Imhoff}}, \bibinfo {author}
		{\bibfnamefont {F.~G.}\ \bibnamefont {Mariam}}, \bibinfo {author}
		{\bibfnamefont {C.~L.}\ \bibnamefont {Morris}}, \bibinfo {author}
		{\bibfnamefont {L.~P.}\ \bibnamefont {Neukirch}}, \bibinfo {author}
		{\bibfnamefont {J.}~\bibnamefont {Perry}}, \bibinfo {author} {\bibfnamefont
			{D.}~\bibnamefont {Poulson}}, \bibinfo {author} {\bibfnamefont
			{R.}~\bibnamefont {Simpson}}, \emph {et~al.},\ }\href
	{https://doi.org/10.1063/1.5011198} {\bibfield  {journal} {\bibinfo
			{journal} {Applied Physics Letters}\ }\textbf {\bibinfo {volume} {112}},\
		\bibinfo {pages} {144103} (\bibinfo {year} {2018})}\BibitemShut {NoStop}%
	\bibitem [{\citenamefont {Xiao}\ \emph {et~al.}(2018)\citenamefont {Xiao},
		\citenamefont {Zhang}, \citenamefont {Cao}, \citenamefont {Yuan},
		\citenamefont {Shen}, \citenamefont {Cheng}, \citenamefont {Zhao},
		\citenamefont {Zong}, \citenamefont {Liu}, \citenamefont {Zhou} \emph
		{et~al.}}]{Xiao_2018}%
	\BibitemOpen
	\bibfield  {author} {\bibinfo {author} {\bibfnamefont {J.}~\bibnamefont
			{Xiao}}, \bibinfo {author} {\bibfnamefont {Z.}~\bibnamefont {Zhang}},
		\bibinfo {author} {\bibfnamefont {S.}~\bibnamefont {Cao}}, \bibinfo {author}
		{\bibfnamefont {P.}~\bibnamefont {Yuan}}, \bibinfo {author} {\bibfnamefont
			{X.}~\bibnamefont {Shen}}, \bibinfo {author} {\bibfnamefont {R.}~\bibnamefont
			{Cheng}}, \bibinfo {author} {\bibfnamefont {Q.}~\bibnamefont {Zhao}},
		\bibinfo {author} {\bibfnamefont {Y.}~\bibnamefont {Zong}}, \bibinfo {author}
		{\bibfnamefont {M.}~\bibnamefont {Liu}}, \bibinfo {author} {\bibfnamefont
			{X.}~\bibnamefont {Zhou}}, \emph {et~al.},\ }\href
	{https://doi.org/10.1088/1674-1056/27/3/035202} {\bibfield  {journal}
		{\bibinfo  {journal} {Chinese Physics B}\ }\textbf {\bibinfo {volume} {27}},\
		\bibinfo {pages} {035202} (\bibinfo {year} {2018})}\BibitemShut {NoStop}%
	\bibitem [{\citenamefont {Zhao}\ \emph {et~al.}(2016)\citenamefont {Zhao},
		\citenamefont {Zhang}, \citenamefont {Gai}, \citenamefont {Du}, \citenamefont
		{Cao}, \citenamefont {Qiu}, \citenamefont {Zhao}, \citenamefont {Cheng},
		\citenamefont {Zhou}, \citenamefont {Ren},\ and\ \citenamefont
		{et~al.}}]{zhao}%
	\BibitemOpen
	\bibfield  {author} {\bibinfo {author} {\bibfnamefont {Y.}~\bibnamefont
			{Zhao}}, \bibinfo {author} {\bibfnamefont {Z.}~\bibnamefont {Zhang}},
		\bibinfo {author} {\bibfnamefont {W.}~\bibnamefont {Gai}}, \bibinfo {author}
		{\bibfnamefont {Y.}~\bibnamefont {Du}}, \bibinfo {author} {\bibfnamefont
			{S.}~\bibnamefont {Cao}}, \bibinfo {author} {\bibfnamefont {J.}~\bibnamefont
			{Qiu}}, \bibinfo {author} {\bibfnamefont {Q.}~\bibnamefont {Zhao}}, \bibinfo
		{author} {\bibfnamefont {R.}~\bibnamefont {Cheng}}, \bibinfo {author}
		{\bibfnamefont {X.}~\bibnamefont {Zhou}}, \bibinfo {author} {\bibfnamefont
			{J.}~\bibnamefont {Ren}},\ and\ \bibinfo {author} {\bibnamefont {et~al.}},\
	}\href {https://doi.org/10.1017/S0263034616000124} {\bibfield  {journal}
		{\bibinfo  {journal} {Laser and Particle Beams}\ }\textbf {\bibinfo {volume}
			{34}},\ \bibinfo {pages} {338–342} (\bibinfo {year} {2016})}\BibitemShut
	{NoStop}%
	\bibitem [{\citenamefont {Xiao}\ \emph {et~al.}(2021)\citenamefont {Xiao},
		\citenamefont {Du}, \citenamefont {Zhang},\ and\ \citenamefont
		{Zhao}}]{article25}%
	\BibitemOpen
	\bibfield  {author} {\bibinfo {author} {\bibfnamefont {J.}~\bibnamefont
			{Xiao}}, \bibinfo {author} {\bibfnamefont {Y.}~\bibnamefont {Du}}, \bibinfo
		{author} {\bibfnamefont {S.}~\bibnamefont {Zhang}},\ and\ \bibinfo {author}
		{\bibfnamefont {Y.}~\bibnamefont {Zhao}},\ }\href
	{https://doi.org/10.1155/2021/6683245} {\bibfield  {journal} {\bibinfo
			{journal} {Laser and Particle Beams}\ }\textbf {\bibinfo {volume} {2021}},\
		\bibinfo {pages} {1} (\bibinfo {year} {2021})}\BibitemShut {NoStop}%
	\bibitem [{\citenamefont {Zhu}\ \emph {et~al.}(2010)\citenamefont {Zhu},
		\citenamefont {Zhang}, \citenamefont {Chen}, \citenamefont {Li},
		\citenamefont {Li}, \citenamefont {Wang}, \citenamefont {Cao}, \citenamefont
		{Sheng},\ and\ \citenamefont {Zhang}}]{doi:10.1063/1.3491994}%
	\BibitemOpen
	\bibfield  {author} {\bibinfo {author} {\bibfnamefont {P.~F.}\ \bibnamefont
			{Zhu}}, \bibinfo {author} {\bibfnamefont {Z.~C.}\ \bibnamefont {Zhang}},
		\bibinfo {author} {\bibfnamefont {L.}~\bibnamefont {Chen}}, \bibinfo {author}
		{\bibfnamefont {R.~Z.}\ \bibnamefont {Li}}, \bibinfo {author} {\bibfnamefont
			{J.~J.}\ \bibnamefont {Li}}, \bibinfo {author} {\bibfnamefont
			{X.}~\bibnamefont {Wang}}, \bibinfo {author} {\bibfnamefont {J.~M.}\
			\bibnamefont {Cao}}, \bibinfo {author} {\bibfnamefont {Z.~M.}\ \bibnamefont
			{Sheng}},\ and\ \bibinfo {author} {\bibfnamefont {J.}~\bibnamefont {Zhang}},\
	}\href {https://doi.org/10.1063/1.3491994} {\bibfield  {journal} {\bibinfo
			{journal} {Review of Scientific Instruments}\ }\textbf {\bibinfo {volume}
			{81}},\ \bibinfo {pages} {103505} (\bibinfo {year} {2010})}\BibitemShut
	{NoStop}%
	\bibitem [{\citenamefont {Li}\ \emph {et~al.}(2010)\citenamefont {Li},
		\citenamefont {Wang}, \citenamefont {Chen}, \citenamefont {Clinite},
		\citenamefont {Mao}, \citenamefont {Zhu}, \citenamefont {Sheng},
		\citenamefont {Zhang},\ and\ \citenamefont {Cao}}]{doi:10.1063/1.3380846}%
	\BibitemOpen
	\bibfield  {author} {\bibinfo {author} {\bibfnamefont {J.}~\bibnamefont
			{Li}}, \bibinfo {author} {\bibfnamefont {X.}~\bibnamefont {Wang}}, \bibinfo
		{author} {\bibfnamefont {Z.}~\bibnamefont {Chen}}, \bibinfo {author}
		{\bibfnamefont {R.}~\bibnamefont {Clinite}}, \bibinfo {author} {\bibfnamefont
			{S.~S.}\ \bibnamefont {Mao}}, \bibinfo {author} {\bibfnamefont
			{P.}~\bibnamefont {Zhu}}, \bibinfo {author} {\bibfnamefont {Z.}~\bibnamefont
			{Sheng}}, \bibinfo {author} {\bibfnamefont {J.}~\bibnamefont {Zhang}},\ and\
		\bibinfo {author} {\bibfnamefont {J.}~\bibnamefont {Cao}},\ }\href
	{https://doi.org/10.1063/1.3380846} {\bibfield  {journal} {\bibinfo
			{journal} {Journal of Applied Physics}\ }\textbf {\bibinfo {volume} {107}},\
		\bibinfo {pages} {083305} (\bibinfo {year} {2010})}\BibitemShut {NoStop}%
	\bibitem [{\citenamefont {Chen}\ \emph {et~al.}(2015)\citenamefont {Chen},
		\citenamefont {Li}, \citenamefont {Chen}, \citenamefont {Zhu}, \citenamefont
		{Liu}, \citenamefont {Cao}, \citenamefont {Sheng},\ and\ \citenamefont
		{Zhang}}]{Chen14479}%
	\BibitemOpen
	\bibfield  {author} {\bibinfo {author} {\bibfnamefont {L.}~\bibnamefont
			{Chen}}, \bibinfo {author} {\bibfnamefont {R.}~\bibnamefont {Li}}, \bibinfo
		{author} {\bibfnamefont {J.}~\bibnamefont {Chen}}, \bibinfo {author}
		{\bibfnamefont {P.}~\bibnamefont {Zhu}}, \bibinfo {author} {\bibfnamefont
			{F.}~\bibnamefont {Liu}}, \bibinfo {author} {\bibfnamefont {J.}~\bibnamefont
			{Cao}}, \bibinfo {author} {\bibfnamefont {Z.}~\bibnamefont {Sheng}},\ and\
		\bibinfo {author} {\bibfnamefont {J.}~\bibnamefont {Zhang}},\ }\href
	{https://doi.org/10.1073/pnas.1518353112} {\bibfield  {journal} {\bibinfo
			{journal} {Proceedings of the National Academy of Sciences}\ }\textbf
		{\bibinfo {volume} {112}},\ \bibinfo {pages} {14479} (\bibinfo {year}
		{2015})}\BibitemShut {NoStop}%
	\bibitem [{\citenamefont {Makino}\ and\ \citenamefont
		{Berz}(2006)}]{MAKINO2006346}%
	\BibitemOpen
	\bibfield  {author} {\bibinfo {author} {\bibfnamefont {K.}~\bibnamefont
			{Makino}}\ and\ \bibinfo {author} {\bibfnamefont {M.}~\bibnamefont {Berz}},\
	}\href {https://doi.org/https://doi.org/10.1016/j.nima.2005.11.109}
	{\bibfield  {journal} {\bibinfo  {journal} {Nuclear Instruments and Methods
				in Physics Research Section A}\ }\textbf {\bibinfo {volume} {558}},\ \bibinfo
		{pages} {346} (\bibinfo {year} {2006})},\ \bibinfo {note} {proceedings of the
		8th International Computational Accelerator Physics Conference}\BibitemShut
	{NoStop}%
	\bibitem [{\citenamefont {Hirayama}\ \emph {et~al.}(2005)\citenamefont
		{Hirayama}, \citenamefont {Namito}, \citenamefont {/KEK}, \citenamefont
		{Bielajew}, \citenamefont {Wilderman}, \citenamefont {U}, \citenamefont
		{Nelson},\ and\ \citenamefont {/SLAC}}]{osti_877459}%
	\BibitemOpen
	\bibfield  {author} {\bibinfo {author} {\bibfnamefont {H.}~\bibnamefont
			{Hirayama}}, \bibinfo {author} {\bibfnamefont {Y.}~\bibnamefont {Namito}},
		\bibinfo {author} {\bibfnamefont {T.}~\bibnamefont {/KEK}}, \bibinfo {author}
		{\bibfnamefont {A.~F.}\ \bibnamefont {Bielajew}}, \bibinfo {author}
		{\bibfnamefont {S.~J.}\ \bibnamefont {Wilderman}}, \bibinfo {author}
		{\bibfnamefont {M.}~\bibnamefont {U}}, \bibinfo {author} {\bibfnamefont
			{W.~R.}\ \bibnamefont {Nelson}},\ and\ \bibinfo {author} {\bibnamefont
			{/SLAC}},\ }\bibfield  {journal} {\bibinfo  {journal} {EGS5}\ }\href
	{https://doi.org/10.2172/877459} {10.2172/877459} (\bibinfo {year}
	{2005})\BibitemShut {NoStop}%
	\bibitem [{\citenamefont {Floettmann}(2017)}]{Astra}%
	\BibitemOpen
	\bibfield  {author} {\bibinfo {author} {\bibfnamefont {K.}~\bibnamefont
			{Floettmann}},\ }\href {https://www.desy.de/~mpyflo/} {\bibfield  {journal}
		{\bibinfo  {journal} {ASTRA}\ } (\bibinfo {year} {2017})}\BibitemShut
	{NoStop}%
\end{thebibliography}
\end{document}